\documentclass{osa-article}

\journal{ome}

\usepackage{graphicx}
\usepackage[T1]{fontenc}
\usepackage{tabularx}
\usepackage{gensymb}
\usepackage{units}
\usepackage{tabu}
\newcommand{\NV}{NV$^{-}$}

\begin{document}

\title{Plasma treatments and photonic nanostructures for shallow nitrogen vacancy centers in diamond}

\author{Mariusz Radtke,\authormark{1} Lara Render,\authormark{1} Richard Nelz,\authormark{1} and Elke Neu\authormark{1,*}}

\address{\authormark{1}Saarland University, Department of Physics, Campus E2.6, 66123 Saarbr\"ucken}

\email{\authormark{*}elkeneu@physik.uni-saarland.de} 



\begin{abstract}
We investigate the influence of plasma treatments, especially a \unit[0]{V}-bias, potentially low damage O$_2$ plasma as well as a biased Ar/SF\textsubscript{6}/O$_2$ plasma on shallow, negative nitrogen vacancy (\NV) centers. We ignite and sustain using our \unit[0]{V}-bias plasma using purely inductive coupling. To this end, we pre-treat surfaces of high purity chemical vapor deposited single-crystal diamond (SCD). Subsequently, we create $\sim$10 nm deep \NV\ centers via implantation and annealing. Onto the annealed SCD surface, we fabricate nanopillar structures that efficiently waveguide the photoluminescence (PL) of shallow \NV. Characterizing single \NV\ inside these nanopillars, we find that the Ar/SF\textsubscript{6}/O$_2$ plasma treatment quenches \NV\ PL even considering that the annealing and cleaning steps following ion implantation remove any surface termination. In contrast, for our \unit[0]{V}-bias as well as biased O$_2$ plasma, we observe stable \NV\ PL and low background fluorescence from the photonic nanostructures.
\end{abstract}

\section{Introduction \label{sec:intro}}
Single-crystal diamond (SCD) with high purity, high crystalline quality and well-controlled surface properties is an enabling material for quantum technologies including quantum sensing \cite{Atature2018,casola2018}. SCD is transparent for visible light and supports light confinement and guiding in photonic nanostructures (refractive index 2.4 visible range, indirect electronic band gap \unit[5.45]{eV}, \cite{Zaitsev2001}). SCD hosts more than 500 optically active point defects termed color centers \cite{Zaitsev2001}. Considering quantum technologies, the currently mostly investigated color center is the negative nitrogen vacancy complex (\NV). \NV\ centers provide controllable, optically readable spins \cite{Gruber1997} and are stable electrical dipoles that emit single photons \cite{Kurtsiefer2000, Babinec2010} or react to optical near-fields \cite{Tisler2013a, nelz2019near}. Their electronic spin degree of freedom renders them sensitive to magnetic fields \cite{Maletinsky2012}, electric fields \cite{Dolde2014}, temperature \cite{Kucsko2013} and strain \cite{Teissier2014} in the diamond matrix. In this context, the key to sensitive measurements with simultaneously nanoscale spatial resolution is to employ individual \NV\ centers placed shallowly (typically $< \unit[10]{nm}$) below SCD surfaces. To enable nanoscale imaging of various samples, it is mandatory to embed the sensing \NV\ in a SCD scanning probe tip \cite{Maletinsky2012, Appel2016}. In addition to enable scanning of a sample, the SCD tip's photonic properties direct the \NV\ photoluminescence (PL) to the collection optics and thus enhance sensitivity \cite{Fuchs2018, Maletinsky2012}. The most prominent candidates for SCD scanning probes are pyramidal tips \cite{Nelz2016, Nicolas2018} and cone-shaped or cylindrical pillars (on platforms) \cite{Maletinsky2012,Appel2016, Fuchs2018}. The latter require sophisticated top-down processes to sculpt the structures from SCD.

Top-down nanofabrication processes reliably create tailored sensing devices potentially enabling highly efficient sensing. However, these processes simultaneously endanger shallow \NV\ centers: To ensure reliable processing, typically smooth (roughness several nanometers), mostly commercially available SCD plates are used. Mechanical polishing of ultra-hard SCD used to obtain these smooth surfaces, however, may create subsurface damage and stress extending several micrometers into SCD \cite{Volpe2009, Naamoun2012}. Consequently, SCD surfaces which are highly suited for nanofabrication might not be optimal to host shallow \NV. To circumvent this challenge, typically several micrometer of SCD are removed using inductively coupled plasma reactive ion etching (ICP-RIE) \cite{Appel2016}. Typical ICP-RIE plasmas, however, have been shown to damage the uppermost few nanometers of the etched SCD \cite{Kato2017}. This damage will also potentially affect shallow \NV\ created under this etched surfaces \cite{Oliveira2015}. Moreover, e.g.\ chlorine gas in the etch plasma is very helpful to keep SCD surfaces smooth during etching \cite{Lee2008,Friel2009} but might attach to the surfaces, potentially de-activating shallow \NV\ \cite{Tao2014}. Low damage plasmas in which the etching species are not accelerated towards the etched SCD (\unit[0]{V}-bias plasmas) have been investigated as a solution to the challenge of ICP-RIE induced damage \cite{Oliveira2015}. Using ICP-RIE here is the method of choice as the density of activated species in the plasma and the acceleration towards the etched substrate can be controlled independently of each other by controlling the power inductively coupled to the plasma (ICP power) and the power coupled capacitively to the plasma (often termed radio frequency (RF) power or platen power) \cite{Hwang2004, Xie2018, Hicks2019}. Previous work \cite{Oliveira2015} employed the \unit[0]{V}-bias plasma after creating shallow \NV. Consequently, the plasma treatment removes SCD containing shallow \NV\ centers and alters the \NV\ density. Uncertainties in the etch rate of the \unit[0]{V}-bias plasma thus transform into uncertainties in \NV\ density rendering the reliable creation of individual \NV\ in nanostructures challenging.  
\begin{figure}
\centering\includegraphics[width=1\linewidth]{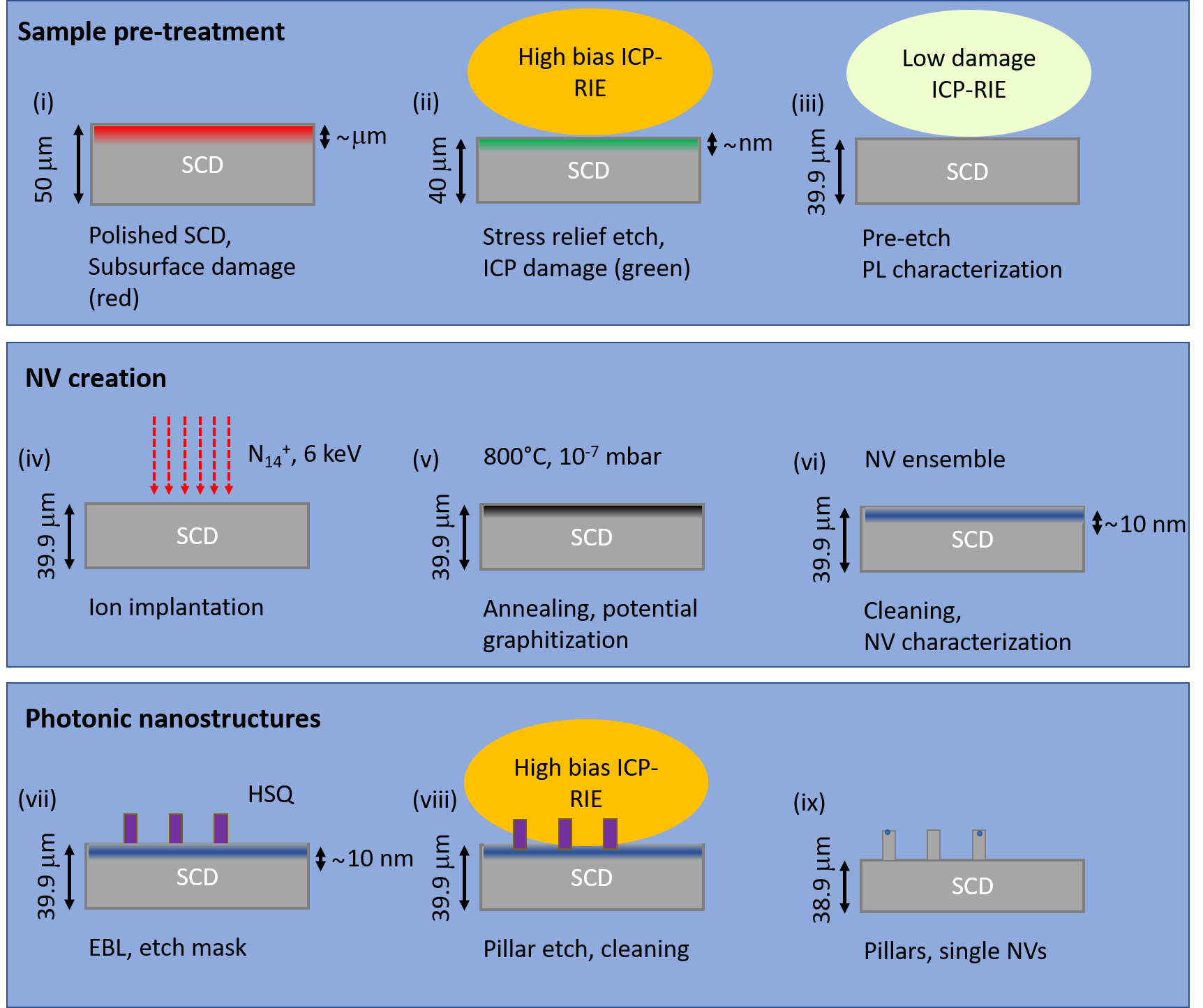}
\caption{Schematic of the employed nanofabrication process. The manuscript is structured into the section according to these schematics. \label{fig:schematics_nanofab}}
\end{figure} 

In this manuscript, we present a process to manufacture photonic nanostructures namely nanopillars with shallow, individual \NV. We, for the first time, include a \unit[0]{V}-bias O$_2$ plasma as pre-treatment for \NV\ creation and SCD nanopillars fabrication thus avoiding changing the \NV\ density by our plasma treatment. Figure \ref{fig:schematics_nanofab} summarizes our nanofabrication process. Following our previous work \cite{Challier2018}, we restrict the used etch gases to non-corrosive, non-toxic gases namely Ar, SF$_6$ and O$_2$.  Starting from polished, commercial SCD plates [Fig.\ \ref{fig:schematics_nanofab}(i)], we use high bias ICP-RIE to remove micrometer thick potentially damaged layers [\textbf{stress relief etch}, Fig.\ \ref{fig:schematics_nanofab}(ii)] while conserving smoothness. We finish this processing step with different plasma treatments, especially a potentially low damage O$_2$ plasma step [\textbf{pre-etch}, Fig.\ \ref{fig:schematics_nanofab}(iii)]. Following this sample pre-treatment, we use established techniques to create \NV\ centers [Figs.\ \ref{fig:schematics_nanofab}(iv)-\ref{fig:schematics_nanofab}(vi)]. We use an optimized process involving electron beam lithography (EBL) and ICP-RIE to structure nanopillars with single \NV\ [Figs.\ \ref{fig:schematics_nanofab}(vii)-\ref{fig:schematics_nanofab}(ix)]. We characterize single \NV\ using PL spectroscopy, PL saturation, PL lifetime, photon correlation measurements as well as optically detected magnetic resonance (ODMR).

\section{Experimental setup and methods}
We use a custom-built confocal scanning microscope (numerical aperture 0.8) to characterize the NV centers' properties in the SCD where confocal filtering is ensured using a single mode fiber. To acquire confocal PL maps, we excite NV centers with a continuous diode-pumped solid-state (DPSS) laser with a wavelength of \unit[532]{nm}. To investigate the charge state of NV centers, we use an additional DPSS laser with a wavelength of \unit[594]{nm}. We detect the PL signal through a \unit[650]{nm} longpass-filter and use highly-efficient photon counters (Excelitas SPCM-AQRH-14, quantum efficiency $\sim$ \unit[70]{\%}) to quantify the PL intensity. In addition, we can send the light to a grating spectrometer (Acton Spectra Pro 2500, Pixis 256OE CCD). We use a tuneable (\unit[450-850]{nm}), pulsed laser (NKT EXW-12, pulse length $\sim$ \unit[50]{ps}) equipped with a filter system (NKT SuperK Varia) and correlation electronics (PicoQuant, PicoHarp 300) to perform time-resolved PL analysis for e.g.\ lifetime measurements. To perform spin manipulation on \NV\ centers in the SCD, we equip the setup with a microwave source (Stanford Research Systems, SG 384) and an amplifier (Mini Circuits, ZHL-42W+) to deliver microwaves through a \unit[20]{$\mu$m} thick gold wire.\\
To perform ICP-RIE, we employ an Oxford Plasmalab 100 ICP RIE system. The plasma lab 100 uses a helical coil ICP configuration. We analyze the plasma composition by means of optical emission spectroscopy (Ocean Optics USB 2000+ coupled by fiber optics to the ICP-RIE system). We employ atomic force microscopy (AFM) measurements (Bruker Fastscan, tapping mode, silicon carbide cantilevers) to deduce the surface roughness of our etched surfaces. To create \NV\ centers, ion implantation has been performed at Augsburg University (Ion-Implantation-System NV 3206, Axcelis Technologies). Samples were annealed at \unit[800]{$^\circ$C} under \unit[$1.5\times10^{-7}$]{mbar} vacuum using a home-built annealing oven incorporating a heater plate (Tectra Boralectric). We employ a cold-cathode scanning electron microscope (SEM) (Hitachi S45000), equipped with RAITH Elphy software to perform electron beam-lithography (EBL).

\section{Sample pre-treatment: stress relief- and pre-etch \label{sec:pretreat}}
In this study, we use commercially available SCD grown by chemical vapor deposition of electronic grade purity (nitrogen, [N$_s^0$] \unit[<5]{ppb} and boron [B]\unit[<1]{ppb}, Element Six, UK). The SCD is polished to form plates with \unit[50]{$\mu$m} thickness by Delaware Diamond Knives, US. Using such thin SCD plates is motivated by our goal to manufacture free-standing SCD scanning probe devices that require thinning the plate to a thickness $<\unit[5]{\mu m}$ \cite{Appel2016}. Here, a thickness of \unit[50]{$\mu$m} is a good compromise between mechanical stability and process time for thinning. To ease handling during nanofabrication, we fix the SCD plate to a silicon carrier using small amounts of crystalbond adhesive. The plates have a roughness of R$_a$=\unit[3]{nm} according to manufacturer specifications and lateral dimensions of \unit[2x4]{mm}. The received diamonds were cleaned by a boiling tri-acid mixture (1:1:1 v/v of H\textsubscript{2}SO\textsubscript{4}, HNO\textsubscript{3}, HClO\textsubscript{4}) and subsequent washing in acetone/isopropanol. We check surface cleanliness in a stereo microscope (50 x magnification) and find no visible residues over the entire SCD surface.

Subsequent to cleaning, we use an etching recipe published previously \cite{Challier2018} to remove the topmost \unit[2.3]{$\mu$m} of our SCD plate. This stress relief etch [see also Fig. \ref{fig:schematics_nanofab}(ii)] is applied uniformly to the full surface of the SCD plate and uses an Ar/SF$_6$/O$_2$ plasma as final step (parameters see Table \ref{tab:etchparameters}). This recipe avoids surface roughening, despite the fact that we are removing potentially damaged SCD material \cite{Challier2018}. As indicated by previous work, adding a fluorine containing etch gas to the process aids in avoiding micromasking on the SCD arising from potential silicon contamination in typical ICP-RIE chambers \cite{Hicks2019}.

\begin{table}
\centering
\begin{tabularx}{\textwidth}{p{1.5cm}p{1cm}p{1cm}p{1cm}p{1cm}p{1cm}p{1cm}p{1cm}}
Plasma&ICP Power (W) & RF Power (W) & DC Bias (V) & Gas Flux (sccm) & Etch Rate (nm/min) & Pressure (mTorr) & rough\-ness (nm) \\
\hline
\unit[435]{V} O\textsubscript{2} & 500 &  200 & 435 &  O\textsubscript{2}: 50& 98 & 11 & 0.7 \\
Ar/SF\textsubscript{6}/O\textsubscript{2}&  700 &  100 &  150 &  O\textsubscript{2} 22   & 87 & 13 & 1 \\
 & & & &  SF\textsubscript{6}:7 & & & \\
& & & & Ar:15& & & \\
\unit[0]{V} O\textsubscript{2} & 550 & 0 & 0 & O\textsubscript{2}: 50 & 8 & 12 &0.7\\
\end{tabularx}
\caption{Plasma parameters for stress relief and pre-etch. The etch rate of the \unit[0]{V}-bias O$_2$ plasma varied between 8 to 10 nm/min and was thus reasonably consistent. \label{tab:etchparameters}}
\end{table}

\begin{figure}
\centering
\includegraphics[width=1\linewidth]{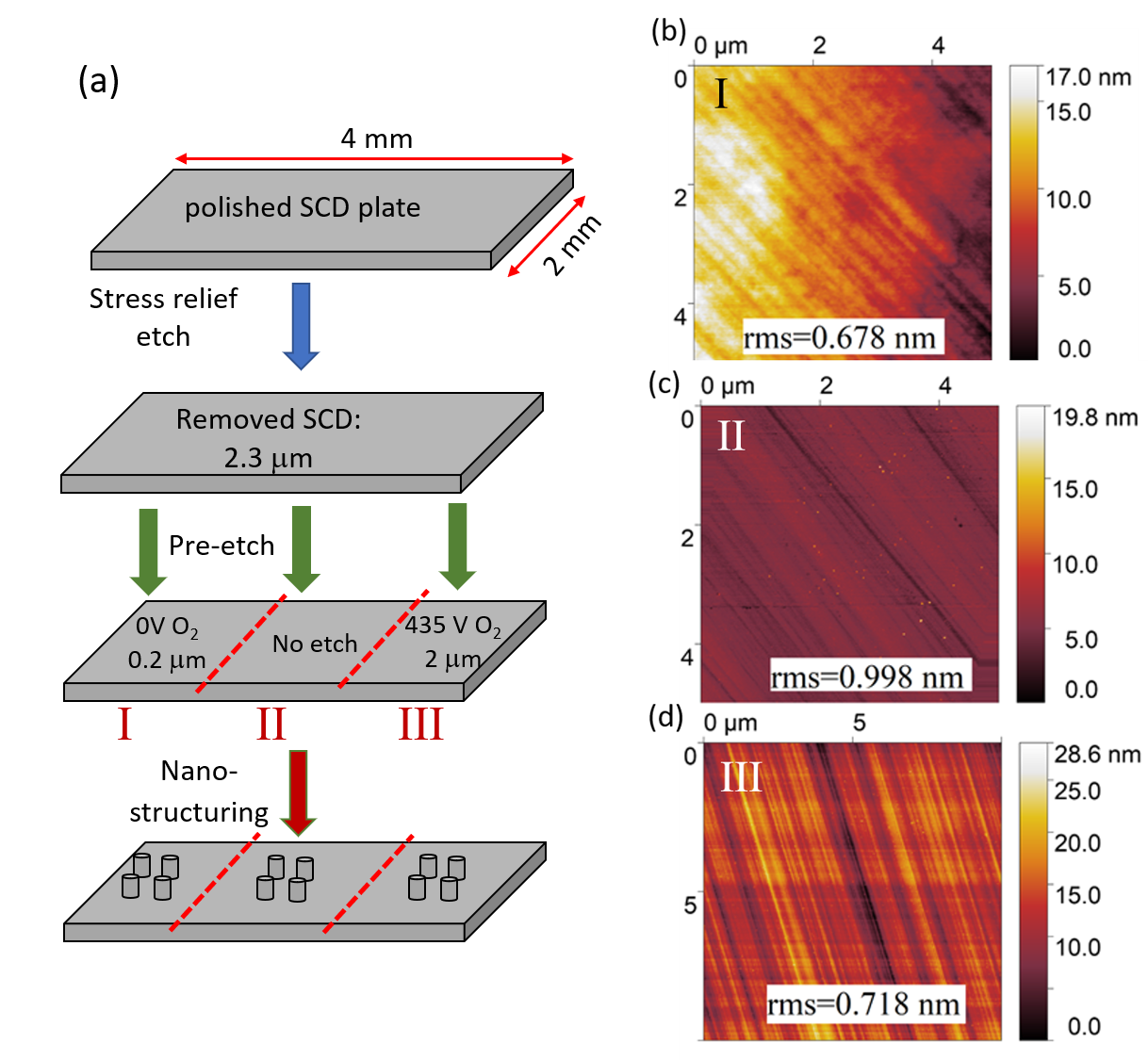}
\caption{(a) Sample layout for the characterization of different plasma treatements. The whole polished SCD plate first undergoes the stress relief etch. Subsequently, we apply different pre-etches to form three distinct areas (plasma type and depth of etch given in the sketch). The ordering of the plasma treatments for the pre-etch step is as follows: We cover areas II and III with a quartz plate (thickness \unit[100]{$\mu$m}) and apply the \unit[0]{V} O$_2$ plasma to area I. By moving the quartz plate and covering areas I and II, the \unit[435]{V} O$_2$ plasma was applied to area III. Subsequently, we fabricate nanopillars in all areas. The pillars are not to scale and for illustration only.  (b)-(d) AFM scans to determine the surface roughness in areas I,II and III. The given roughness is the rms value obtained from \unit[5x5]{$\mu$m$^2$} tapping mode scans. \label{fig:layout0Vchem}}
\end{figure}
While the stress relief etch procedure reliably removes even tens of micrometers of SCD keeping the surface smooth \cite{Challier2018}, highly energetic etch species in the plasma potentially damage the first nanometers of the SCD \cite{Kato2017}. In this study, we thus systematically investigate the effect of  additional plasma steps, termed pre-etch plasmas, which follow the stress relief etch and by which we terminate the removal of material [see Fig.\ \ref{fig:schematics_nanofab}(iii)]. We aim to compare different pre-etch treatments applied to the same SCD plate. We thus rule out that sample quality changes influence our findings. To this end, we protect not to be etched areas using a \unit[100]{$\mu$m} thin quartz plate and create three areas in the SCD plate [see Fig.\ \ref{fig:layout0Vchem}(a)]. In two areas, we apply additional pre-etch plasmas (\unit[0]{V}-bias O$_2$ area I, \unit[435]{V}-bias O$_2$ area III) prior to creating shallow \NV\ centers (see Sec.\ \ref{sec:NV_creation}). In contrast, area II is not treated by a pre-etch plasma and was consequently only etched by the Ar/SF$_6$/O$_2$ plasma of the stress relief etch.

Table \ref{tab:etchparameters} summarizes the parameters of the employed etch plasmas. We note that a larger variety of \unit[0]{V}-bias plasmas has been tested but only the \unit[0]{V}-bias O$_2$ plasma enabled etching without excess surface roughening (see Appendix \ref{sec:suppl:0V}). We also take great care to ignite the \unit[0]{V}-bias plasma without bias building up in its starting phase by careful adjustments of capacitors in the reactor and without using capacitively coupled radio frequency power.
As discernible from Table \ref{tab:etchparameters}, the biased plasmas reach etch rates of almost \unit[100]{nm/min}, while the \unit[0]{V}-bias O$_2$ plasma's etch rate is an order of magnitude lower. As indicated above, \unit[2.3]{$\mu$m} of SCD were removed in the stress-relief etch (and consequently in area II). In area I, we additionally remove $\sim$\unit[200]{nm} using the \unit[0]{V}-bias O$_2$ plasma. In area III, we additionally remove \unit[2]{$\mu$m} using the \unit[435]{V}-bias O$_2$ plasma.

After the pre-etches, we again clean the SCD sample as described above and measure the surface roughness in areas I, II and III using AFM. The results of these measurements are depicted in Figs.\ \ref{fig:layout0Vchem}(b)-\ref{fig:layout0Vchem}(d) and summarized in Table \ref{tab:etchparameters}. For all areas, we find an rms roughness below \unit[1]{nm}, consequently all regions are usable for the fabrication of nanopillars with shallow \NV. Smooth surfaces are mandatory for high spatial resolution, scanning probe sensing using \NV\ as any surface roughness will transform into uncontrolled stand-off distances between \NV\ sensor and sample under investigation. 

\begin{figure}
\centering
\includegraphics[width=1\linewidth]{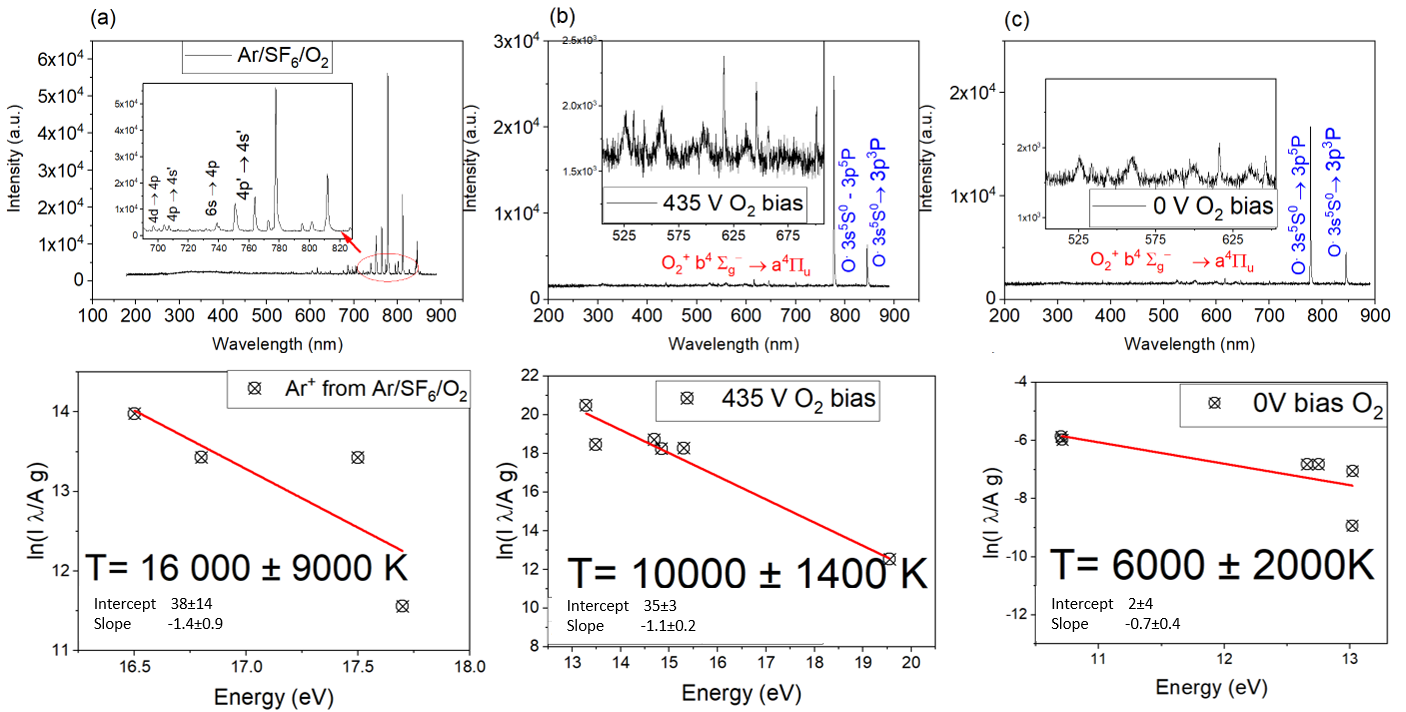}
\caption{Optical emission spectra of (a) Ar/SF\textsubscript{6}/O\textsubscript{2}, (b) O$_2$ 435 V and (c) O\textsubscript{2} 0 V (350 W ICP) plasmas with corresponding Boltzmann plots used to extract electron temperatures $T_{el}$. \label{Boltzmann}}
\end{figure}

In the following, we discuss the etch mechanisms underlying the different plasma processes. While all surfaces show low roughness, remarkably the surface etched with the \unit[0]{V}-bias O$_2$ plasma shows the lowest roughness. \unit[0]{V} plasmas have been found to etch isotropically \cite{Khanaliloo2015, Khanaliloo2015a} via chemical etching effects. This situation might induce micromasking due to e.g.\ residual impurities/dust on the SCD surface that cannot be removed by this soft plasma or lead to preferential etching of defective areas \cite{Hicks2019}. However, we do not observe pronounced micromasking in our work. We use optical emission spectroscopy to compare the composition of reactive species in our plasmas (see Fig.\ \ref{Boltzmann}). Observing the molecular transitions depicted in the insets of Figs.\ \ref{Boltzmann}(b) and \ref{Boltzmann}(c), we analyze the concentration of ionized oxygen O$_2^+$. In a biased O$_2$ plasma, O$_2^+$ is the main etching species as the charged ions are effectively accelerated towards the SCD sample and will induce physical etching \cite{PhysRevLett.116.025001, bazaka_oxygen_2018}. We find a 50 times higher concentration of O$_2^+$ in the \unit[435]{V}-bias O$_2$ plasma compared to the \unit[0]{V}-bias O$_2$ plasma (see Fig.\ \ref{Boltzmann}). In contrast, the etch rate of the \unit[0]{V}-bias O$_2$ plasma is only roughly one order of magnitude lower than for the \unit[435]{V} O$_2$ plasma indicating an additional etching species in the \unit[0]{V}-bias plasma. We suggest this additional etching species to be (atomic) oxygen radicals. Observing the transitions 3s$^{5}$S$^{0}\rightarrow$ 3p$^{5}$P and 3s$^{5}$S$^{0}\rightarrow$ 3p$^{3}$P in the oxygen radical system, we find almost the same concentration of oxygen radicals in the \unit[0]{V}-bias and \unit[435]{V}-bias plasmas. For the \unit[0]{V}-bias plasma, chemical etching based on the oxygen radicals is dominant which leads, as expected \cite{SARANGAN2016149}, to much lower etch rates (see Table \ref{tab:etchparameters}). We expect chemical etching induced by non-accelerated radicals to introduce less damage to the SCD surface compared to the etching via ionized species accelerated under a DC bias. However, chemical etching is isotropic \cite{Khanaliloo2015, Khanaliloo2015a} and therefore not applicable for nanostructures that require steep sidewalls e.g.\ nanopillars.

To additionally characterize the employed O$_2$ plasmas, we investigate their electron temperature $T_{el}$. 
Electrons in the plasma will collide with the molecules/atoms of the etch gas exciting the gas. Consequently, higher electron densities transfer into a higher population of excited states in molecules/atoms. In the absence of a DC bias, the electron temperature can be an upper estimate for the thermal energy of the etching species in the plasma. It should be noted that the electrons might not fully transfer their thermal energy to the etching species but are still a clear indication of the plasma density \cite{Zhang2016a}. In contrast in the case of a biased plasma, the energy of the etching species will be mainly governed by the DC bias \cite{Xie2018}. We use the measured optical emission spectra of the plasmas to extract $T_{el}$. First, we consider that
\begin{equation}\label{eq:boltz}
\ln{\frac{I \lambda}{g_kA}}=- \frac{E_k}{kT_{el}} +C.
\end{equation} 
Here, $\lambda$ is the wavelength and $I$ the measured intensity of each transition. $g_k$ is the statistical weight of the upper excited level, $k$ the Boltzmann constant, $E_k$ the energy of the excited upper level, $A$ the Einstein coefficient for the respective transition and $C$ is the constant of integration. We plot $\ln{\frac{I \lambda}{g_kA}}$ as a function of $E_k$ and thus the slope measured is equal to $-\frac{1}{kT_{el}}$ and $T_{el}$ can be extracted. The electronic transitions taken into account to determine $T_{el}$ are marked in Fig.\ \ref{Boltzmann}. 

We find $T_{el}$=\unit[6000 $\pm$ 2000]{K} for the \unit[0]{V} O$_2$ plasma, $T_{el}$=\unit[10000 $\pm$ 1400]{K} for the \unit[435]{V} O$_2$ plasma and $T_{el}$=\unit[16000 $\pm$ 9000]{K} for our Ar/SF$\textsubscript{6}$/O$\textsubscript{2}$ plasma. For the O$_2$ plasma especially the following transitions were chosen: $^{5}\textrm{P}_{_{3}} \rightarrow ^{3}\textrm{S}_{_{2}}^{^{0}}$ (544.04 nm),$^{5}\textrm{P}_{_{1}} \rightarrow ^{5}\textrm{S}_{_{2}}^{^{0}}$ (543.06 nm),$^{3}\textrm{P}_{_{1}} \rightarrow ^{5}\textrm{S}_{_{2}}^{^{0}}$ (646.18 nm), $^{3}\textrm{D}_{_{2}} \rightarrow ^{3}\textrm{P}_{_{2}}^{^{0}}$(615.83 nm), $^{3}\textrm{D}_{_{1}} \rightarrow ^{3}\textrm{D}_{_{0}}^{^{0}}$(599.45 nm),$^{3}\textrm{D}_{_{1}} \rightarrow ^{3}\textrm{D}_{_{0}}^{^{1}}$ (601.19 nm). These values clearly indicate the higher density of the biased plasmas.

After determining the surface roughness of our SCD samples, we also analyze the PL originating from the etched surfaces. Broadband PL from processed SCD surfaces is detrimental in experiments with individual \NV\ as it will lead to low signal-to-background ratios for single color center observation. Here, we find a spatially homogeneous background PL (detected at wavelengths > \unit[650]{nm}, excited using 532 nm laser light) in the order of \unit[30]{kcps/mW} for both O$_2$-based pre-etches (areas I and III). The observed PL is stable under continuous laser illumination. In addition, we show that the background PL arises purely from the SCD surface as no PL is observed upon focusing the laser deeply (> 5 $\mu$m) into the SCD material. Thus, in contrast to previous work \cite{Oliveira2015} that compared a biased Ar/O$_2$ plasma with a \unit[0]{V}-O$_2$ plasma, we do not find a reduction of PL background for the \unit[0]{V}-O$_2$ plasma. We note that the observed background PL from these surfaces is in the same order of magnitude as the PL from single \NV\ centers. We find similar background levels after creating shallow \NV\ centers (see section \ref{sec:NV_creation}). However, the fabrication of photonic nanostructures allows us to efficiently reduce the influence of background PL (see section \ref{sec:photonic_nano}).

In stark contrast, the SCD surface exposed to the Ar/SF$_6$/O$_2$ plasma (area II) shows an intense, fast-bleaching background PL. The background that bleaches within $\sim$ \unit[100]{ms} is in the order of more than \unit[7.5]{Mcps/mW}. For examples of the PL maps recorded on these surfaces, see Appendix Fig.\ \ref{fig:Suppl:PreChar}. For diamond films, typically broadband PL in the red-spectral range is observed and attributed to highly defective material (sp$^2$ inclusions, disorder, amorphous carbon \cite{Bergman1994}). Significant broadband PL in the red spectral range thus indicates the presence of damaged material close to the surface which could not be removed using the acid clean. Interestingly, this background PL is removed after NV creation steps (see Sec. \ref{sec:NV_creation}). We suspect that this is connected to the annealing process which in conjunction with acid cleaning aids in removing damaged layers at SCD surfaces \cite{Kato2017}. 

\section{Creation of NV centers \label{sec:NV_creation}}
\begin{figure}
\centering
\includegraphics[width=1\textwidth]{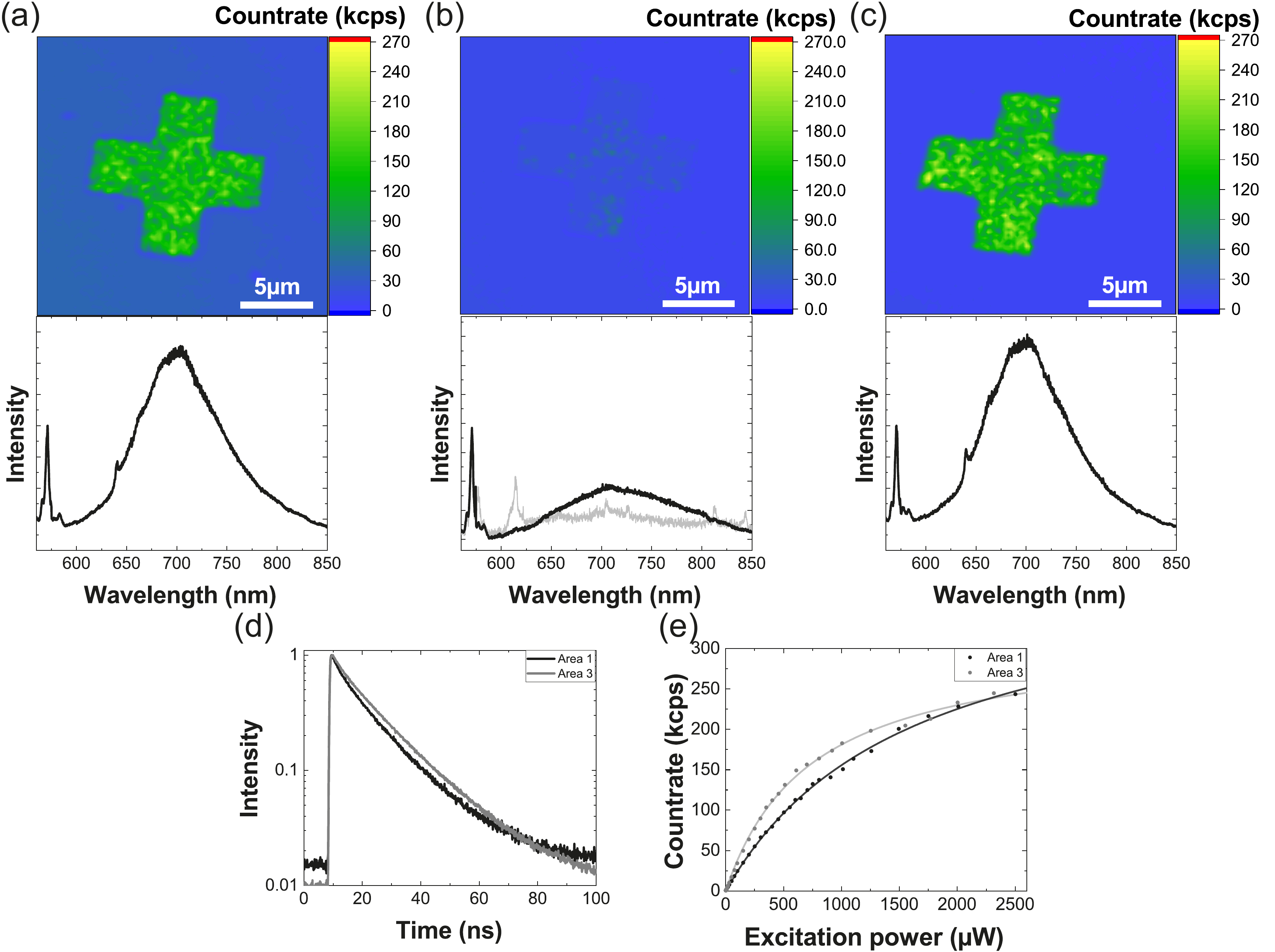}
\caption{\NV\ center ensembles in areas exposed to different pre-etches: (a)-(c) show PL maps and the corresponding PL spectra. It is clearly visible that in the area exposed to the Ar/SF$_6$/O$_2$ stress-relief etch only [Area II, part (b)] we do not observe any \NV\ PL. The measured spectrum only slightly differs from the background PL (light gray). Both areas treated with pure O$_2$ plasmas [Area I (unbiased):(a); Area III (biased):(c)] clearly show \NV\ PL with a minor NV$^0$ signal. The peak at \unit[574]{nm} is the first-order Raman line of SCD. Due to experimental constraints (laser linewidth, spectrometer resolution) the Raman line here does not allow to deduce the level of strain in the SCD material. (d) Excited state lifetime $\tau_{NV^-}$ of the \NV\ ensemble as recorded in areas I and III. Both ensembles show similar average $\tau_{NV^-}$ \unit[17(1)]{ns} in area I and \unit[16(1)]{ns} in area III. $\tau_{NV^-}$ is thus prolonged compared to bulk SCD due to the \NV's proximity to the surface \cite{Nelz2019}. In addition, the measurements clearly show an absence of quenching. (e) Saturation behavior of the \NV\ ensembles area I and III. Discussion see text. \label{fig:marker}}
\end{figure}

Subsequently, we implant the whole surface of our SCD plate using ${}^{14}$N$^+$ ions (Energy \unit[6]{keV}, $\unit[2 \times 10^{11}]{\nicefrac{\text{ions}}{\text{cm}^2}}$). We create \NV\ centers via annealing at \unit[800]{$^\circ$C} for 4 hours. We note that directly after this procedure, we typically do not observe any \NV\ PL and the emission spectrum of the SCD surface shows non-diamond Raman lines. Re-cleaning the sample in tri-acid mixture activates the \NV\ PL which we attribute to the removal of non-diamond, graphitic phases from the SCD surface. We note that alternatively thin layers of \NV\ centers can also be introduced using $\delta$-doping, reducing crystal damage due to implantation \cite{deOliveira2016}. \\
We first characterize the influence of the plasma treatments on \NV\ centers by investigating the implanted \NV\ ensembles. To obtain illustrative results in the different areas, we investigate \NV\ ensembles in micrometer-sized marker structures (crosses) that we obtained during nanofabrication (see Sec.\ \ref{sec:photonic_nano}). These structures do not significantly alter the collection efficiency for \NV\ PL or the properties of \NV\ ensembles due to their size in the micrometer range. However, they give us the possibility to clearly distinguish \NV\ PL from background PL of the SCD surface as well as to clearly compare the different areas. First, we record PL maps of crosses and the corresponding PL spectra in areas I, II and III [see Figs.\ \ref{fig:marker}(a)-\ref{fig:marker}(c)].
As clearly discernible from Fig.\ \ref{fig:marker}, in areas I and III, the \NV\ ensemble has similar PL brightness and the recorded PL spectra clearly reveal the spectral features of \NV\ PL. In contrast, in area II treated using the Ar/SF$_6$/O$_2$ etch, we do not find any \NV\ PL nor any evidence for neutral NV$^0$ centers. We note that we do not expect this effect to result from surface termination: First, fluorine terminated surfaces are supposed to stabilize \NV\ \cite{Osterkamp2015}. Second, any surface termination would have most probably been removed during vacuum annealing and replaced by oxygen functional groups during the acid cleaning. The complete absence of an \NV\ fingerprint supports the assumption of highly-damaged layers which prevented the creation of \NV\ centers or strongly quench their PL.

In addition, we measure the excited state lifetime $\tau_{NV^-}$ [see Fig.\ \ref{fig:marker}(d)] of the \NV\ ensemble in areas I and III. We obtain comparable results for both areas with $\tau_{NV^-}$=\unit[17(1)]{ns} for area I and $\tau_{NV^-}$=\unit[16(1)]{ns} for area III, respectively. Here, $\tau_{NV^-}$ is longer than the bulk lifetime (\unit[12]{ns}) due to the \NV's proximity to the surface \cite{Nelz2019}. However, we exclude quenching which would reduce $\tau_{NV^-}$.

To further characterize the influence of the plasma-treatments onto the brightness of the \NV\ ensemble, we measure PL saturation curves in areas I and III. Here, we note that we investigate a \NV\ ensemble with a low density of $\sim$7 \NV\ centers statistically distributed in the laser focus of our confocal microscope (density estimate see Sec.\ \ref{sec:photonic_nano}). Consequently, the observed saturation behavior and the parameters extracted from it are only estimates as few \NV\ centers experience different laser intensities as well as different collection efficiencies. We find saturation powers $P_{sat}$ which are in the same order of magnitude (\unit[1.60(5)]{mW} area I, \unit[0.74(3)]{mW} area III) as well as comparable background PL rates ($\sim$\unit[60]{kcps/mW} area I, $\sim$\unit[40]{kcps/mW} area III) and count rates ($I_\infty = $\unit[407(7)]{kcps} area I, $I_\infty = $\unit[315(5)]{kcps} area III) in both O$_2$-etched areas. We also note that the background PL level estimated here agrees with the level measured in area I and III before creating \NV\ centers (see Sec.\ \ref{sec:pretreat}).

By applying microwave-driven spin manipulation to the NV ensembles in areas I and III, we are able to measure their coherence time $T_2 < $\unit[10]{$\mu$s} limited mainly by the proximity of the \NV\ centers to the surface which agrees well with other measurements of shallow \NV\ centers under tri-acid cleaned surfaces \cite{sangtawesin2018}. In contrast to previous work \cite{Oliveira2015}, we do not observe an enhancement of $T_2$ as a result of the \unit[0]{V}-bias O$_2$ plasma treatment. We consequently conclude that $T_2$ for our \NV\ centers is mainly governed by noise due to surface termination and we are not yet able to reveal positive effects of the \unit[0]{V}-bias O$_2$ plasma treatment. As we do not observe significant strain-induced splittings of ODMR resonance, we conclude that our \NV\ centers are not experiencing significant strain.    
\section{\NV\ in photonic nanostructures \label{sec:photonic_nano}}
To asses the stability as well as brightness of individual \NV\ centers created under the \unit[0]{V}-bias O$_2$ plasma treated surface, we fabricate SCD nanopillars in the shape of truncated cones. We aim for diameters of the pillars' top facet that contains the \NV\ center in the range of 200 nm. To this end, we use a refined process compared to previously published processes \cite{Babinec2010,Appel2016,Neu2014}, details of the process are published elsewhere \cite{Radtkenanofab2019}. The steps in the process are:
\begin{itemize}
    \item Deposition of a silicon adhesion layer onto SCD.
    \item Spin coating of hydrogen silsesquioxane (HSQ)-based negative tone EBL resist (Fox 16, Dow Corning).
    \item EBL to create pillar masks, development of resist.
    \item ICP-RIE removal of adhesion layer and subsequent etching of SCD to form pillars, details see Appendix \ref{sec:pillarfab} and Ref.\ \cite{Radtkenanofab2019} 
    \item residual mask stripping in buffered oxide etch and sample cleaning (acid clean).
\end{itemize}

\begin{figure}
\centering\includegraphics[width=10cm]{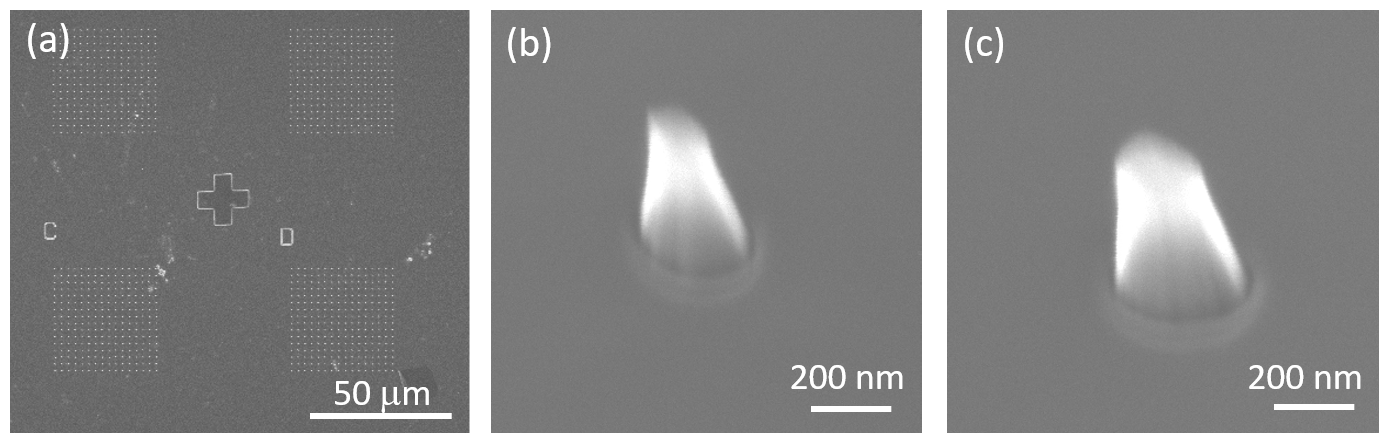}
\caption{Scanning electron microscope images of SCD nanopillars. (a) overview of large pillar field, (b) pillars with top diameter of \unit[120]{nm} (c) pillars with top diameter of \unit[180]{nm} Images recorded using a Jeol arm200 microscope at \unit[20]{kV} acceleration voltage with no conductive layer applied. \label{fig:pillars}} 
\end{figure}

To demonstrate reliable nanofabrication on all plasma treated surfaces, we generate various patterns of nanopillars on the SCD sample. Using our optimized fabrication process, we reliably create large fields of pillars with high yield [see Fig.\ \ref{fig:pillars}(a)]. Due to the novelty of our \unit[0]{V}-bias O$_2$ plasma pre-treatment, we focus on area I. For consistency, we also check pillars with individual NV centers in area III, confirming stable, bright PL in accordance with previous work \cite{Appel2016}. From various pillars fields written in area I using different EBL doses (\unit[1.96-2.52]{$\nicefrac{m C}{\text{cm}^2}$}) and etched using different plasmas (Ar, Ar/O$_2$ pure O$_2$, for details see Appendix \ref{sec:pillarfab}), we focus on two fields in which we were straightforwardly able to identify single \NV\ in the pillars. The tapered pillars written with an EBL dose of \unit[2.2]{$\nicefrac{m \text{C}}{\text{cm}^2}$} show top diameters of \unit[120]{nm} and \unit[180]{nm} [see Figs.\ \ref{fig:pillars}(b) and \ref{fig:pillars}(c)].

The investigated pillars have been etched using a pure O$_2$ plasma (\unit[500]{W} ICP power, \unit[50]{W} radio frequency (RF) power, \unit[50]{sccm} O$_2$, pressure \unit[1.5]{Pa}). We found O$_2$ plasmas to be most reliable for etching our SCD pillars and we obtain an etch rate of \unit[65]{$\nicefrac{\text{nm}}{\text{min}}$}. We observe smooth surfaces in-between the pillars without any indication of micromasking. In contrast to the O$_2$ plasma, mask erosion and micromasking was strong using a pure argon plasma due to excess sputtering of material. We note that the investigated pillars have a stronger taper angle than we typically aim for \cite{Fuchs2018}. The novel fabrication process used here includes a SF$_6$-based plasma to remove the silicon adhesion layer which could potentially attack HSQ masks \cite{Radtkenanofab2019}. We however find that this plasma etches our thin (\unit[25]{nm}) silicon adhesion layer 20 times faster than the HSQ mask and should not cause mask erosion and strong tapering \cite{Radtkenanofab2019}. We consequently suspect an additional effect causing the tapering. One possibility might be a low thickness of the HSQ layer as the mask structures were not imaged using SEM prior to etching the pillars to avoid contamination. Also strong mask faceting might have occurred that is known to cause tapered sidewalls \cite{Toros2018}. 

\begin{figure}
\centering
\includegraphics[width=1\textwidth]{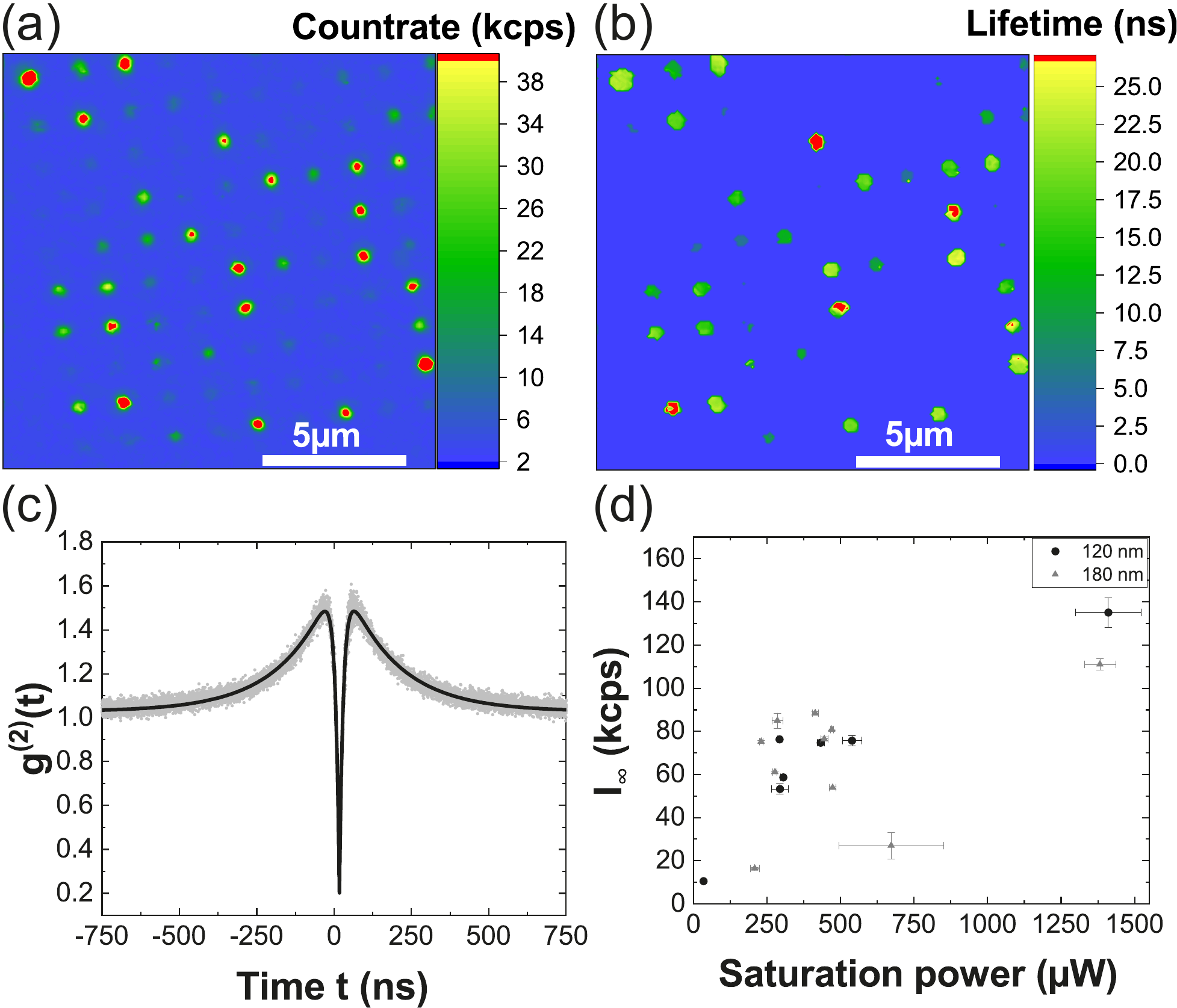}
\caption{(a) and (b) Fluorescence maps and lifetime $\tau_{NV^-}$ maps of a pillar field (top diameter: \unit[180]{nm}, length: $\sim$\unit[460]{nm}) in area I (\unit[0]{V} O$_2$ plasma). (a) shows the measured PL rate under pulsed excitation at \unit[530-540]{nm} with a pulse energy of \unit[15]{nJ} and a repetition rate of \unit[8]{MHz} while (b) shows the corresponding lifetime map. It is nicely visible that bright pillars have a consistent lifetime corresponding to $\tau_{NV^-}$ measured for the ensemble in Sec.\ \ref{sec:NV_creation}. (c) Exemplary second order correlation measurement $g^{(2)}$ of a \NV\ center in the pillars clearly showing single-photon emission with a reasonable signal/background ratio. (d) Summary of PL saturation of several pillars containing single \NV\ centers in the fields with top diameters of \unit[120]{nm} and \unit[180]{nm}. While most measured \NV\ have comparable saturation powers, the \unit[180]{nm}-pillars offer slightly higher countrates compared to the \unit[120]{nm}-pillars as expected from their photonic properties \cite{Fuchs2018}. However, these results indicate that even \unit[120]{nm}-pillars act as waveguides, potentially improving AFM operation of \NV-based scanning probe devices. \label{fig:structures}}
\end{figure}

To characterize \NV\ centers in our photonic nanostructures, we first measure PL and lifetime maps of the pillar fields [see Fig.\ \ref{fig:structures}(a) and \ref{fig:structures}(b)]. Here, we clearly observe a comparable lifetime for most of the pillars containing \NV\ centers. In addition, the average lifetime of $\tau_{NV^-}$=\unit[17(2)]{ns} agrees with $\tau_{NV^-}$ we find for the \NV\ ensemble in Sec.\ \ref{sec:NV_creation}. By measuring second order correlation functions $g^{(2)}$ [see Fig.\ \ref{fig:structures}(c)], we estimate a \NV\ probability per pillar. We then use the pillars' top diameter from SEM images (Fig.\ \ref{fig:pillars}) to calculate the \NV\ density. Subsequently, we compare this with the implantation dose and extract an implantation yield of \unit[0.01]{\NV/implanted N} which is in good agreement with literature \cite{Pezzagna2010,Appel2016}.
To estimate the PL enhancement in our nanostructures, we perform several PL saturation measurements on pillars containing single \NV\ centers. These measurements are summarized in Fig.\ \ref{fig:structures}(d). Here, we compare \NV\ centers in pillars with a length of $\sim$ \unit[460]{nm} and diameters of \unit[120]{nm} and \unit[180]{nm}, respectively. As discernible from Fig.\ \ref{fig:structures}(d), most measured \NV\ centers show similar saturation power. The \unit[180]{nm}-pillars offer slightly higher PL rates as the \unit[120]{nm}-pillars. Simulations suggest a much stronger decrease in PL rate from pillars thinner than \unit[180]{nm} \cite{Fuchs2018}. However, the here observed brightness of single \NV\ PL also indicates a sufficient usability of \unit[120]{nm}-pillars for sensing applications. For most scanning probe based sensing applications, it would be advantageous to use pillars with a small top diameter, as a thinner scanning probe tip will be able to follow the topography of a sample more closely and keep the \NV\ center closer to the surface under investigation e.g.\ when scanning over a step in the sample \cite{Appel2016}. We note that we were not able to deduce \NV\ and NV$^0$ probabilities for our NV centers \cite{Aslam2013,bluvstein2018, Shields2015} due to inconclusive results (see Appendix \ref{sec:Suppl:ChargeState}).\\

\section{Conclusions} 
In the presented study, we investigate a chemically etching, pure ICP discharge, \unit[0]{V}-bias O$_2$ plasma as a pre-etch for the fabrication of SCD photonic structures. We compare this treatment to a biased O$_2$ plasma as well as a Ar/O$_2$/SF$_6$ plasma. We analyze plasma compositions and etch rates. While the biased plasmas have etch rates exceeding \unit[100]{nm/min}, the \unit[0]{V}-bias plasma shows an order of magnitude lower etch rate. While etching in the biased O$_2$ plasma is dominated by oxygen ions, oxygen radicals dominate in the \unit[0]{V}-bias case. Remarkably, we only succeed in creating shallow \NV\ under the O$_2$ treated surfaces, while all \NV\ centers are deactivated under the Ar/O$_2$/SF$_6$ treated surface. We investigate single \NV\ centers in the nanopillars with \unit[180]{nm} as well as \unit[120]{nm} top diameter. We find stable, bright emission from single \NV\ rendering our structures suitable for nanoscale sensing using single \NV. We have shown that \unit[0]{V}-bias plasmas can be integrated into the fabrication of SCD photonic nanostructures as a pre-treatment, opening the route towards low damage treatments of SCD surfaces within SCD nanodevice fabrication.

\section{Appendix}
\subsection{Testing of \unit[0]{V} plasmas and sample layout \label{sec:suppl:0V}}
\begin{figure}[h!]
\centering\includegraphics[width=0.7\linewidth]{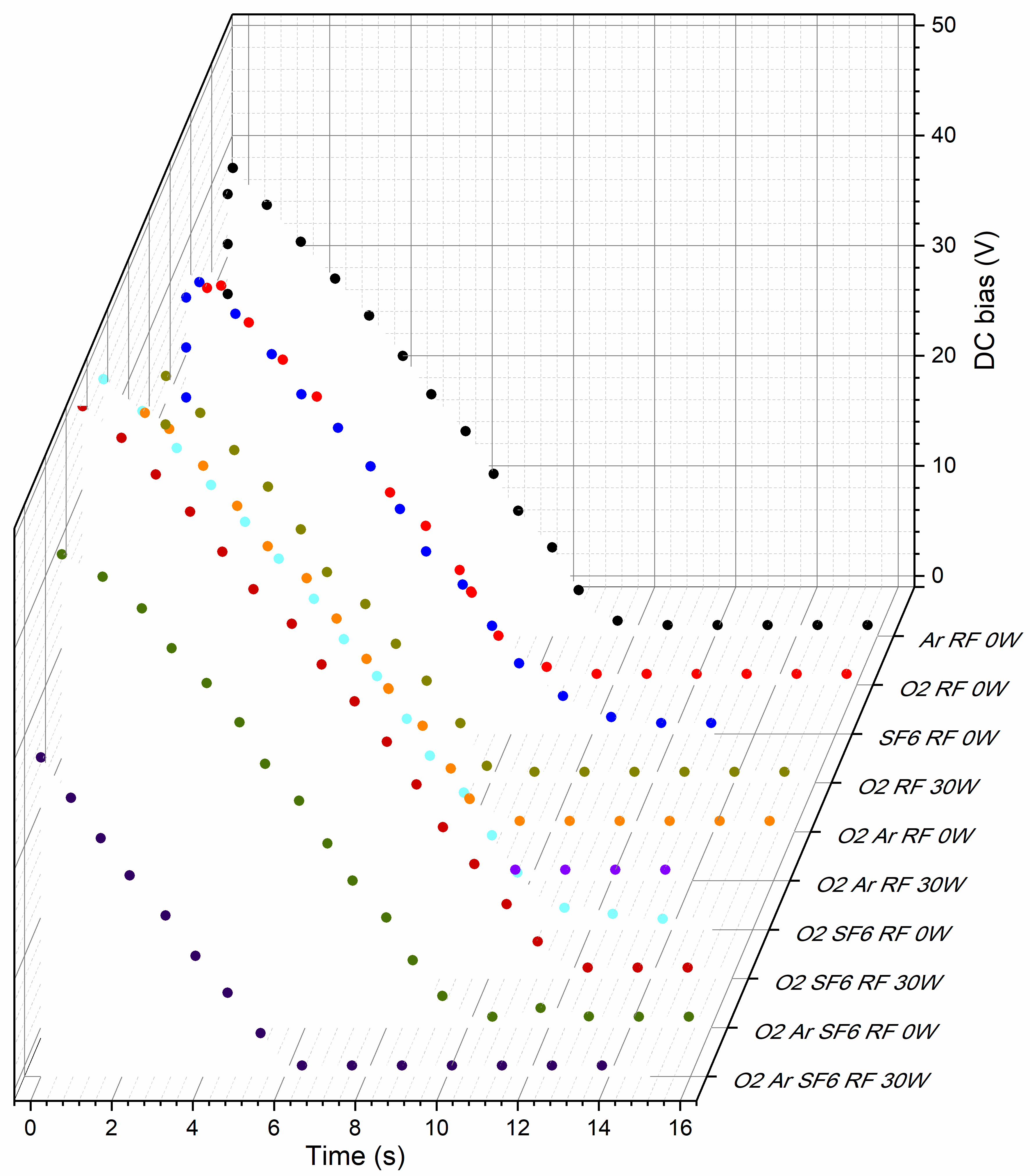}
\caption{Bias drop in time in generation of \unit[0]{V}-bias plasmas using O\textsubscript{2}, Ar, SF\textsubscript{6} and their mixtures.  \label{fig:various0Vplasmas}}
\end{figure}

To check the applicability of \unit[0]{V}-bias plasmas to pre-etch our SCD surfaces, we run different plasmas. 
Figure \ref{fig:various0Vplasmas} summarizes the tested \unit[0]{V} plasmas. We select only the \unit[0]{V} O$_2$ plasma for the main study due to the observed large surface roughening of \unit[0]{V} Ar/SF$_6$ and Ar plasmas (rms $\geq$ 10 nm). We furthermore expect no significant etch rate in a \unit[0]{V} SF$_6$ plasma and do thus also not use this plasma.  Figure \ref{fig:various0Vplasmas} shows the bias drop in time for each plasma after the ignition. This drop could be eliminated by optimizing the matching circuit parameter conditions- which required a run of pre-tests before actual etching prior to each process. The bias drop was found to be comparable in time for each plasma and was dependent on the chamber pressure and gas flux. For the studies presented in the manuscript, we used the optimized parameters for the matching circuit and consequently avoid any buildup of bias. 

\subsection{Characterization of the pre-etched areas}\label{sec:Suppl:Prechar}
As described in the main text, we characterize the three pre-etched areas after etching and cleaning before implanting nitrogen ions into the SCD.\\
\begin{figure}[h!]
\centering
\includegraphics[width=1\textwidth]{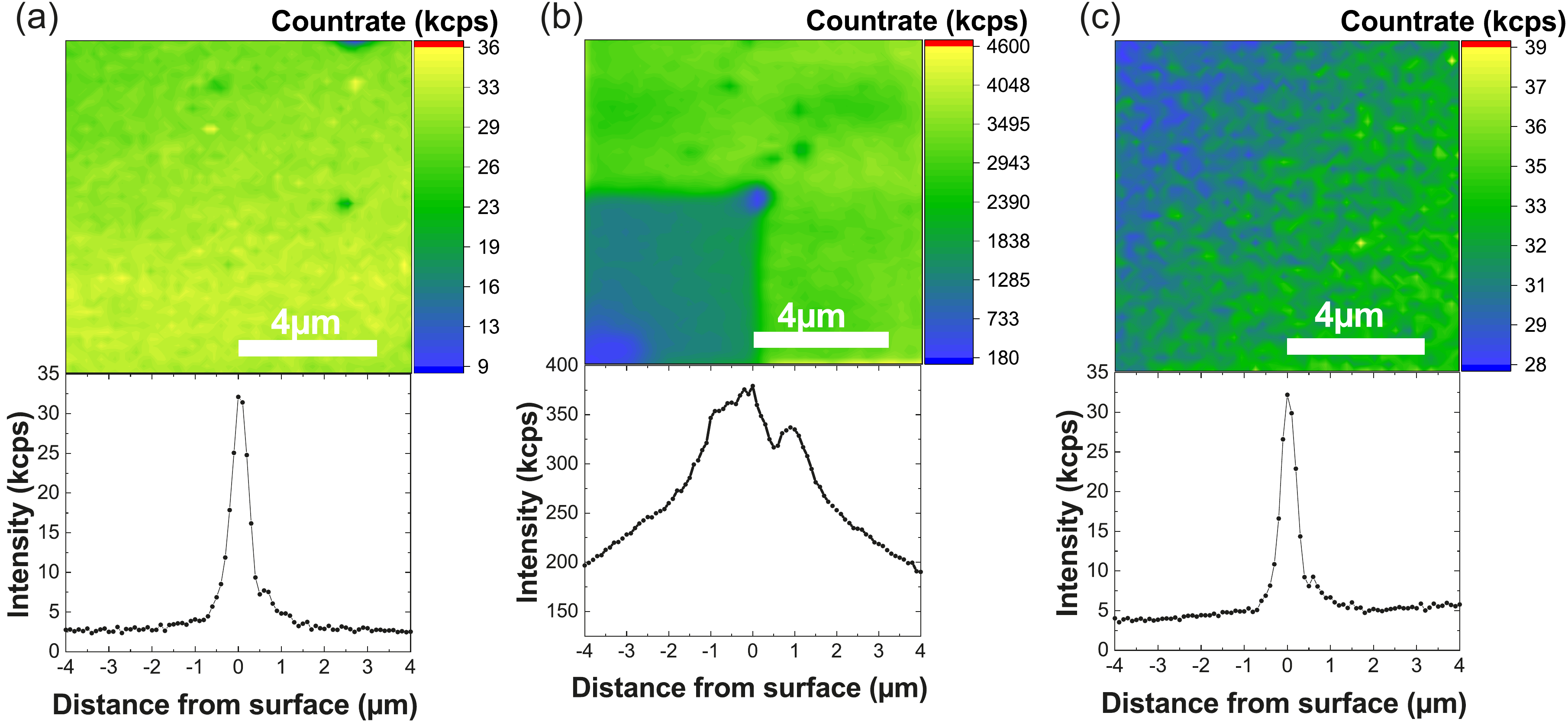}
\caption{Optical characterization of the three pre-etched areas: (a) unbiased O$_2$, (b) Ar/SF$_6$/O$_2$, (c) biased O$_2$. For further information, see text.}
\label{fig:Suppl:PreChar}
\end{figure}
Here, we see a similar behavior of the two oxygen etched areas leading to a spatially homogeneous background PL of \unit[30]{kcps/mW} [see Figs.\ \ref{fig:Suppl:PreChar}(a) and \ref{fig:Suppl:PreChar}(c), area I and III]. In contrast, the Ar/SF$_6$/O$_2$-etched area II shows a different behavior: Here, we clearly see a high background PL in the order of \unit[7.5]{Mcps/mW} which is rapidly bleaching within roughly \unit[100]{ms}. This is clearly visible in Fig.\ \ref{fig:Suppl:PreChar}(b). Here, we first measure a smaller PL map and afterwards a larger PL map around it. This second PL map clearly shows the bleached square of the first measurement. By further focusing the laser onto a single point (as can be seen in the lower left and upper right corner of the smaller square) we are able to further bleach the PL to PL rates comparable with the oxygen-etched areas. In addition, we scan the focus of our confocal laser microscope through the SCD surface (z-scans). Z-scans in all areas clearly show that the background PL originates only from the SCD surface. Note, that z-scans in area II [Fig.\ \ref{fig:Suppl:PreChar}(b)] do not show a clear peak due to the bleaching behavior which is triggered by focusing onto the surface.

\subsection{Details on pillar fabrication \label{sec:pillarfab}} 
\subsubsection{Technical details on electron beam lithography}
Diamond plates immobilized on a silicon chip and spin coated with negative tone, HSQ-based resist FOX-16 were inserted into cold-cathode SEM (Hitachi S45000), equipped with RAITH Elphy software. The electron beam is calibrated on a Si chip containing nanoparticles and the writing process is performed at \unit[30]{kV} acceleration voltage and 20 $\mu$A extracting current.  The beam current was measured always prior the lithography session and applied as correction for the writing dose in the software. The z-distance was kept constant at 15.3 nm for 400x400 $\mu$m fields. The writing process occurs at 100x magnification. The writing mode for long structures is kept longitudinal and for small structures concentric. The area dose is kept as 1400 $\mu$As/cm\textsuperscript{2}. The dose was established to be  2.24 mC/cm\textsuperscript{2} for the pillar structures. Varying the dose in the electron beam lithography, influences the shape of the resulting nanopillars as well as their diameter.
\subsubsection{Details on pillar etching}
Areas I, II and III were divided into three sub-parts each. We fabricated various pillars masks using a variety of EBL parameters out of HSQ that were treated with various oxygen and argon plasmas to elucidate the effects of DC bias, RF/ICP power and gas flux on the geometry of the pillars.\\

 In order to remove the adhesive layer used at the SCD/HSQ interface, a specialized SF$_6$-based etching plasma was applied prior to etching the SCD pillars for 5 seconds. 
\begin{figure}[h!]
\centering\includegraphics[width=10cm] {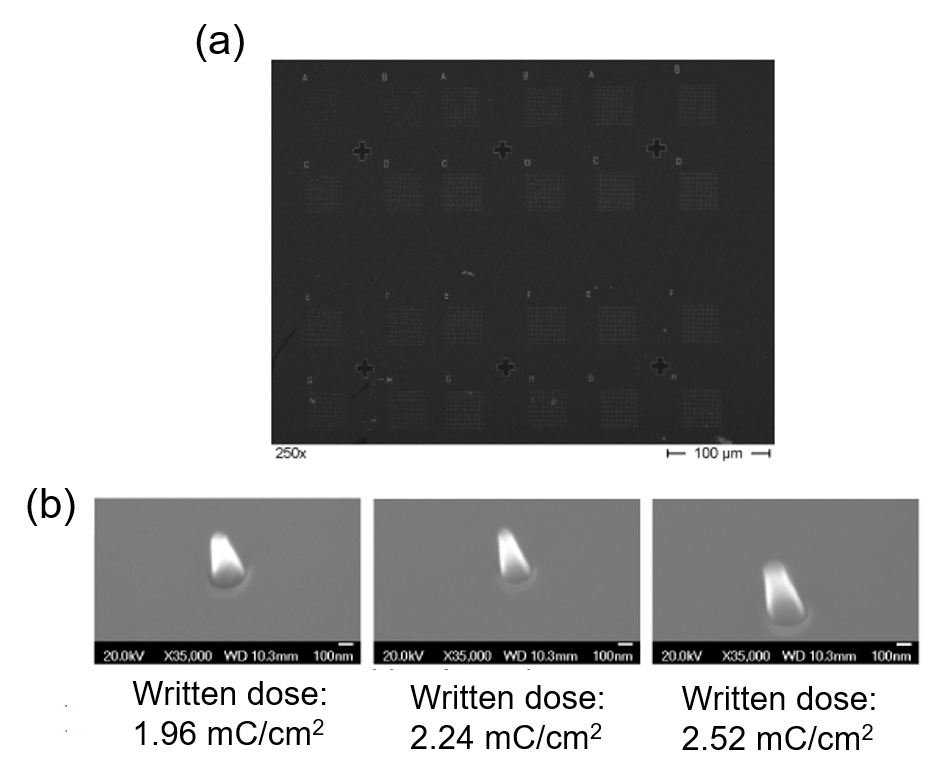}\label{pillars}
\caption{SEM images of SCD nanopillars: (a) overview image of  large pillar fields etched into the diamond by O\textsubscript{2} plasma (b) High magnification scanning electron microscope (Inset B, Jeol arm200, 20 kV acceleration voltage, no conductive layer applied). \label{fig:pillarssuppl}} 
\end{figure}
  Figure \ref{fig:pillarssuppl} shows detailed scanning electron microscope images of etched SCD nanostructures. Fig.\ \ref{fig:pillarssuppl}(a) shows large pillar fields as an indication of the reliability of the employed nanofabrication process. Fig.\ \ref{fig:pillarssuppl}(b) gives examples of nanopillars written with different EBL doses and a nominal mask diameter of \unit[180]{nm}. As clearly discernible from Fig.\ \ref{fig56}, increasing the dose increases the effective mask size and consequently the top diameter of the pillar. All SEM images were acquired at 20 kV acceleration voltage without the use of conductive layer.  The mask erosion and resulting micromasking was especially visible with use of pure Ar plasma probably due top sputtering effects and we observed the generation of needles and damaged surface. This effect was lowered, while mixing argon with other gases (SF\textsubscript{6}, O\textsubscript{2}) and totally removed by using only biased oxygen plasma for the etching. We achieved the best control over the pillar shape using a O\textsubscript{2} plasma at 750 W ICP power and 50 W RF power, 90 $\pm$5 nm etch rate, 2.2 mC/cm\textsuperscript{2} writing dose for the FOX-16 mask (compare Table \ref{tab:pillaretches}). 
\begin{table}
\centering
\begin{tabularx}{\textwidth}{lllllll}
Plasma & DC bias & RF power (W) & ICP Power (W)  & etch rate (nm/min)  \\
\hline
O\textsubscript{2}/Ar & 191 & 50& 500 &--\\
O\textsubscript{2}    & 170 & 50& 500& 65 \\
Ar & 149 & 50 & 500& 20\\
O\textsubscript{2}/Ar & 157& 50& 750 & 65\\
O\textsubscript{2}    & 144 & 50& 750& 90\\
Ar& 119 & 50& 750& 10 
\end{tabularx}
\caption{Parameters for different etch plasmas. Due to excess mask erosion etch rates of the pure Ar plasmas have to be considered coarse estimates.  \label{tab:pillaretches} }
\end{table}

\begin{figure}[h!]
\centering
\includegraphics[width=0.75\textwidth]{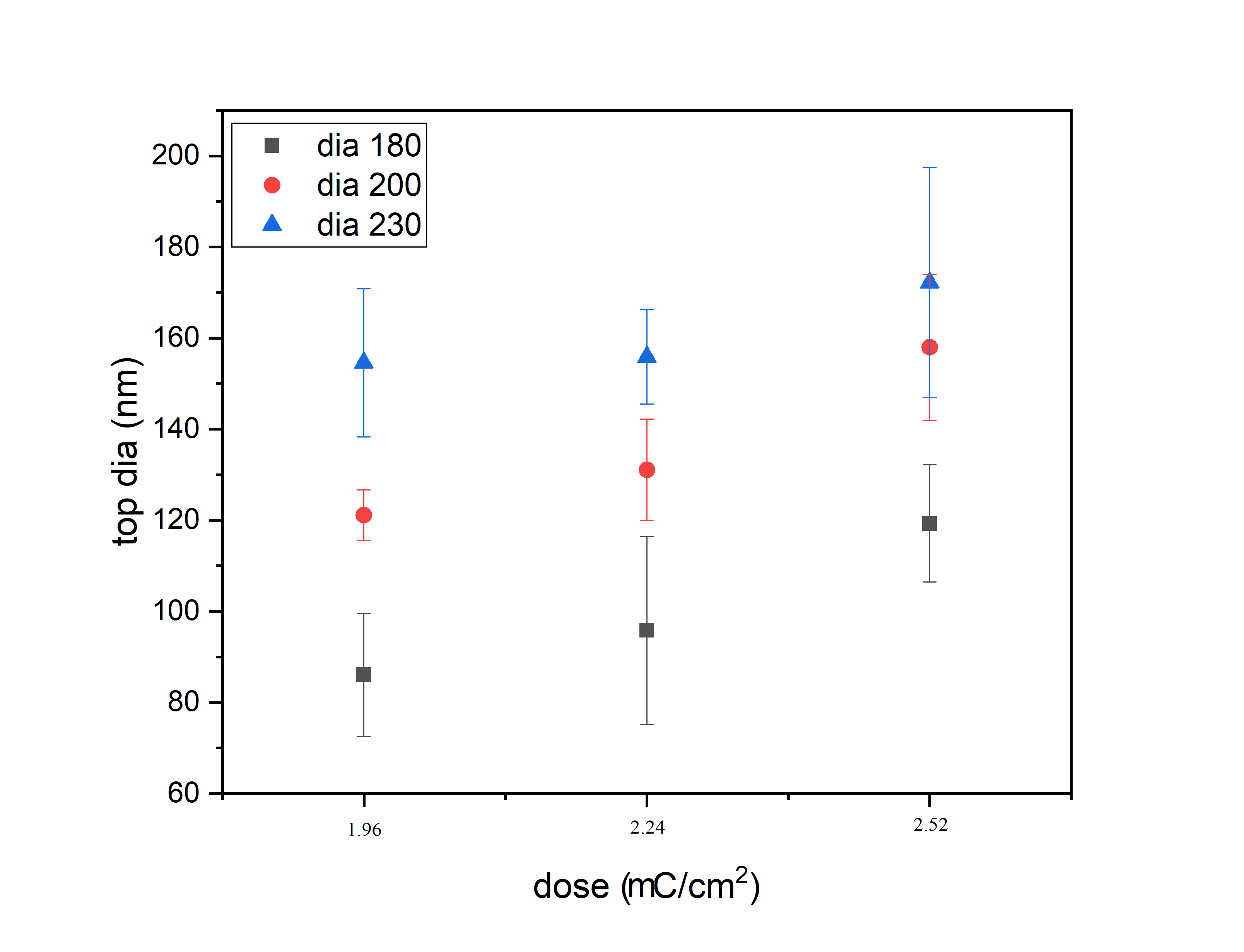}
\caption{Correlation between the EBL dose and the resulting top diameter of the nanopillars. The investigated pillars have been etched using a pure O\textsubscript{2} plasma with 500 W ICP power and 50 W RF power.}
\label{fig56}
\end{figure}

\subsection{Charge state detection}\label{sec:Suppl:ChargeState}
\begin{figure}[h!]
\centering
\includegraphics[width=1\textwidth]{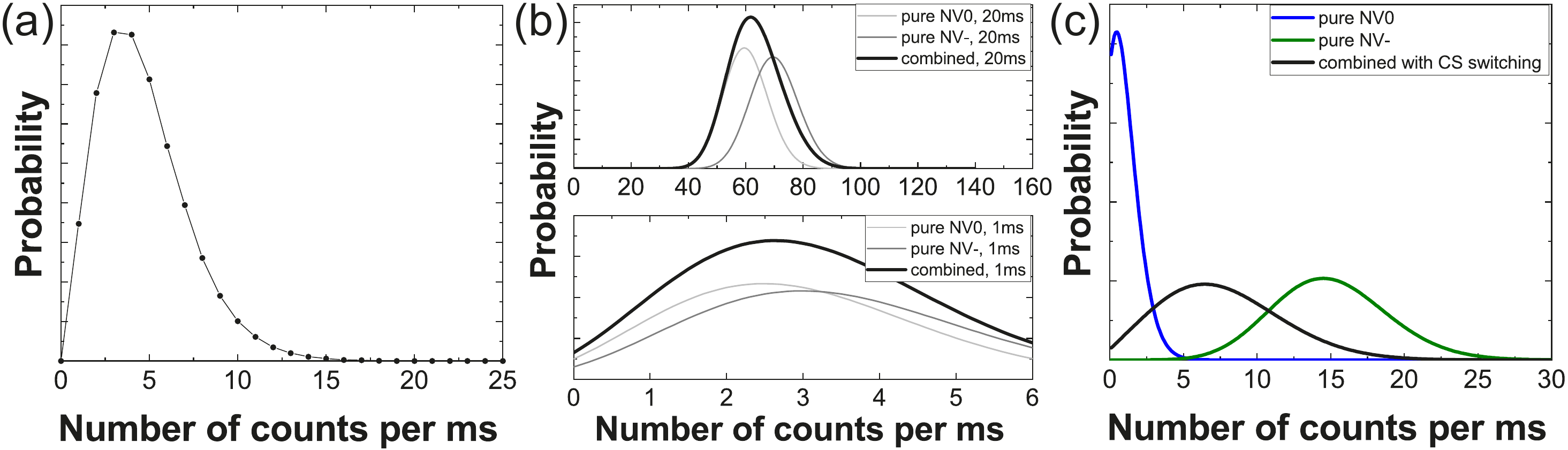}
\caption{(a) Exemplary measurement of the charge state distribution we obtain on the NV centers in the nanostructures. Here, we use the following parameters: \unit[532]{nm} laser: \unit[1]{ms} at \unit[100]{$\mu$W}, \unit[594]{nm}-laser: \unit[1]{ms} at \unit[15]{$\mu$W}, read-out time: \unit[1]{ms}. While using excitation powers ranging from \unit[1]{$\mu$W} to \unit[250]{$\mu$W} for both lasers and integration times up to \unit[25]{ms}, the mean values of the distributions are just shifting without a significant change of the distribution. (b-c) Calculated photon statistics for illustrating the behavior for special cases described in the text.}
\label{fig:Suppl:ChargeState}
\end{figure}

To further study the influence of the different pre-etches on single NV$^-$ centers, we investigate the switching between NV$^-$ and NV$^0$ for our centers according to \cite{Aslam2013,bluvstein2018, Shields2015}. Typically, this measurement shows two Poisson distributions whereas the distribution at low count rates corresponds to NV$^0$ and the second one to NV$^-$ PL. We here find one distribution broader than usual [see Fig.\ \ref{fig:Suppl:ChargeState}(a)]. This could be explained by the following: First, both distributions are joint hindering a clear distinction of both distributions [see Fig.\ \ref{fig:Suppl:ChargeState}(b)]. This could be mainly arising from enhanced background PL. This enhanced background might arise from residuals that were created during repeated cleaning of our sample before the charge state investigations. Second, these residuals could also cause a change of the electrical environment of the NV centers potentially increasing the charge state switching rates  drastically. When this rate exceeds 1/read-out time, both distributions start to merge and eventually become one distribution centered around the mean PL of both charge states [see Fig.\ \ref{fig:Suppl:ChargeState}(c)]. Due to the limited amount of photons we detect under weak excitation with the \unit[594]{nm} laser, we are not able to optimize our read-out time for fast charge state switching rates.\\
To further explain the behavior of the charge state detection measurement, we first have to briefly introduce the photon statistics model used to describe the process. Here, we follow the derivation of \cite{bluvstein2018,Shields2015}.\\ 
By allowing the charge state of the NV center to switch during the measurement, the resulting photon probability changes from two pure Poissonian distributions to a more complex distribution which involves a sum over an infinite number of Poisson distributions weighted by the probability for switching the charge state \cite{bluvstein2018,Shields2015}. This leads to a complex behavior:\\ 
First, we describe the case of the Poissonian distributions lying closely together. This belongs to the case of a strong background PL or a weak NV$^-$ PL as shown in Fig.\ \ref{fig:Suppl:ChargeState}(b). Both panels in Fig.\ \ref{fig:Suppl:ChargeState}(b) show the same mean count-rate of the background and NV$^-$, respectively, with different read-out times (\unit[1]{ms} in the lower panel, \unit[20]{ms} in the upper). Here, it is clearly visible that even by significantly increasing the read out time of the charge state detection it is not possible to distinguish two distributions when the difference of background PL and NV$^-$ PL is too low. Note, we use a negligible charge state switching rate for this calculation.\\
Second, we describe the case of a fast switching between the charge states of the NV center compared to the read-out time. This is shown in Fig.\ \ref{fig:Suppl:ChargeState}(c). Here, we have two clearly distinguishable distributions for the pure NV charge states (blue for NV$^0$ and olive for NV$^-$). By calculating the complex photon statistics with ionization rates exceeding 1/read-out time, we get the black curve. This shows that even  distinguishable mean photon rates can result in an indistinguishable photon distribution in the case of fast charge state switching. 
\section*{Funding}
We acknowledge funding via a NanoMatFutur grant of the German Ministry of Education and Research (FKZ13N13547).

\section*{Acknowledgments}
We would like to acknowledge Dr.\ Matthias Schreck and Wolfgang Br{\"u}ckner (Augsburg University, Germany) for performing nitrogen ion implantations, Dr.\ Ing.\ Sandra Wolff (TU Kaiserslautern, Germany) for help with electron beam evaporation, J{\"o}rg Schmauch (Saarland University, Germany) for his help with acquiring high-quality SEM images and Dr. Rene Hensel (INM, Germany) for granting access to the ICP RIE reactor. We acknowledge Michel Challier for his assistance. 

\section*{Disclosures}We note that the nanofabrication method applied in this study is filed for a patent, application number: EP19198772.6. The authors declare no conflicts of interest.

\bibliography{sample,Literaturverzeichnisaktuell}

\begin{thebibliography}{10}
\newcommand{\enquote}[1]{``#1''}

\bibitem{Atature2018}
M.~Atat{\"u}re, D.~Englund, N.~Vamivakas, S.-Y. Lee, and J.~Wrachtrup,
  \enquote{Material platforms for spin-based photonic quantum technologies,}
  {\protect\JournalTitle{Nature Reviews Materials}} \textbf{3}, 38 (2018).

\bibitem{casola2018}
F.~Casola, T.~van~der Sar, and A.~Yacoby, \enquote{Probing condensed matter
  physics with magnetometry based on nitrogen-vacancy centres in diamond,}
  {\protect\JournalTitle{Nature Reviews Materials}} \textbf{3}, 17088 (2018).

\bibitem{Zaitsev2001}
A.~Zaitsev, \emph{{Optical Properties of Diamond: A Data Handbook}} (Springer,
  2001).

\bibitem{Gruber1997}
A.~Gruber, A.~Dr\"abenstedt, C.~Tietz, L.~Fleury, J.~Wrachtrup, and C.~von
  Borczyskowski, \enquote{{Scanning confocal optical microscopy and magnetic
  resonance on single defect centers},} {\protect\JournalTitle{Science}}
  \textbf{{276}}, 2012--2014 ({1997}).

\bibitem{Kurtsiefer2000}
C.~Kurtsiefer, S.~Mayer, P.~Zarda, and H.~Weinfurter, \enquote{Stable
  solid-state source of single photons,} {\protect\JournalTitle{Phys. Rev.
  Lett.}} \textbf{85}, 290--293 (2000).

\bibitem{Babinec2010}
T.~Babinec, B.~Hausmann, M.~Khan, Y.~Zhang, J.~Maze, P.~Hemmer, and M.~Loncar,
  \enquote{{A diamond nanowire single-photon source},}
  {\protect\JournalTitle{Nature Nanotech.}} \textbf{5}, 195--199 (2010).

\bibitem{Tisler2013a}
J.~Tisler, T.~Oeckinghaus, R.~J. St\"ohr, R.~Kolesov, R.~Reuter, F.~Reinhard,
  and J.~Wrachtrup, \enquote{Single defect center scanning near-field optical
  microscopy on graphene,} {\protect\JournalTitle{Nano Lett.}} \textbf{13},
  3152--3156 (2013).

\bibitem{nelz2019near}
R.~Nelz, M.~Radtke, A.~Slablab, M.~Kianinia, C.~Li, Z.-Q. Xu, C.~Bradac,
  I.~Aharonovich, and E.~Neu, \enquote{Near-field energy transfer between a
  luminescent 2d material and color centers in diamond,}
  {\protect\JournalTitle{Advanced Quantum Technologies}} \textbf{DOI:
  10.1002/qute.201900088} (2019).

\bibitem{Maletinsky2012}
P.~Maletinsky, S.~Hong, M.~Grinolds, B.~Hausmann, M.~Lukin, R.~Walsworth,
  M.~Loncar, and A.~Yacoby, \enquote{A robust scanning diamond sensor for
  nanoscale imaging with single nitrogen-vacancy centres,}
  {\protect\JournalTitle{Nat. Nanotechnol.}} \textbf{7}, 320--324 (2012).

\bibitem{Dolde2014}
F.~Dolde, M.~W. Doherty, J.~Michl, I.~Jakobi, B.~Naydenov, S.~Pezzagna,
  J.~Meijer, P.~Neumann, F.~Jelezko, N.~B. Manson, and J.~Wrachtrup,
  \enquote{Nanoscale detection of a single fundamental charge in ambient
  conditions using the nv- center in diamond,} {\protect\JournalTitle{Phys.
  Rev. Lett.}} \textbf{112}, 097603 (2014).

\bibitem{Kucsko2013}
G.~Kucsko, P.~Maurer, N.~Yao, M.~Kubo, H.~Noh, P.~Lo, H.~Park, and M.~Lukin,
  \enquote{Nanometre-scale thermometry in a living cell,}
  {\protect\JournalTitle{Nature}} \textbf{500}, 54--58 (2013).

\bibitem{Teissier2014}
J.~Teissier, A.~Barfuss, P.~Appel, E.~Neu, and P.~Maletinsky, \enquote{Strain
  coupling of a nitrogen-vacancy center spin to a diamond mechanical
  oscillator,} {\protect\JournalTitle{Phys. Rev. Lett.}} \textbf{113}, 020503
  (2014).

\bibitem{Appel2016}
P.~Appel, E.~Neu, M.~Ganzhorn, A.~Barfuss, M.~Batzer, M.~Gratz, A.~Tsch{\"o}pe,
  and P.~Maletinsky, \enquote{Fabrication of all diamond scanning probes for
  nanoscale magnetometry,} {\protect\JournalTitle{Review of Scientific
  Instruments}} \textbf{87}, 063703 (2016).

\bibitem{Fuchs2018}
P.~Fuchs, M.~Challier, and E.~Neu, \enquote{Optimized single-crystal diamond
  scanning probes for high sensitivity magnetometry,}
  {\protect\JournalTitle{New Journal of Physics}} \textbf{20}, 125001 (2018).

\bibitem{Nelz2016}
R.~Nelz, P.~Fuchs, O.~Opaluch, S.~Sonusen, N.~Savenko, V.~Podgursky, and
  E.~Neu, \enquote{Color center fluorescence and spin manipulation in single
  crystal, pyramidal diamond tips,} {\protect\JournalTitle{Applied Physics
  Letters}} \textbf{109}, 193105 (2016).

\bibitem{Nicolas2018}
L.~Nicolas, T.~Delord, P.~Huillery, E.~Neu, and G.~H{\'e}tet, \enquote{Diamond
  nano-pyramids with narrow linewidth siv centers for quantum technologies,}
  {\protect\JournalTitle{AIP Advances}} \textbf{8}, 065102 (2018).

\bibitem{Volpe2009}
P.-N. Volpe, P.~Muret, F.~Omnes, J.~Achard, F.~Silva, O.~Brinza, and
  A.~Gicquel, \enquote{Defect analysis and excitons diffusion in undoped
  homoepitaxial diamond films after polishing and oxygen plasma etching,}
  {\protect\JournalTitle{Diamond Relat. Mater.}} \textbf{18}, 1205 -- 1210
  (2009).

\bibitem{Naamoun2012}
M.~Naamoun, A.~Tallaire, F.~Silva, J.~Achard, P.~Doppelt, and A.~Gicquel,
  \enquote{Etch-pit formation mechanism induced on hpht and cvd diamond single
  crystals by h2/o2 plasma etching treatment,} {\protect\JournalTitle{physica
  status solidi (a)}} \textbf{209}, 1715--1720 (2012).

\bibitem{Kato2017}
Y.~Kato, H.~Kawashima, T.~Makino, M.~Ogura, A.~Traor{\'e}, N.~Ozawa, and
  S.~Yamasaki, \enquote{Estimation of inductively coupled plasma etching damage
  of boron-doped diamond using x-ray photoelectron spectroscopy,}
  {\protect\JournalTitle{physica status solidi (a)}} \textbf{214}, 1700233
  (2017).

\bibitem{Oliveira2015}
F.~F. de~Oliveira, S.~A. Momenzadeh, Y.~Wang, M.~Konuma, M.~Markham, A.~M.
  Edmonds, A.~Denisenko, and J.~Wrachtrup, \enquote{{Effect of low-damage
  inductively coupled plasma on shallow nitrogen-vacancy centers in diamond},}
  {\protect\JournalTitle{{APPLIED PHYSICS LETTERS}}} \textbf{{107}}, 073107
  ({2015}).

\bibitem{Lee2008}
C.~Lee, E.~Gu, M.~Dawson, I.~Friel, and G.~Scarsbrook, \enquote{Etching and
  micro-optics fabrication in diamond using chlorine-based inductively-coupled
  plasma,} {\protect\JournalTitle{Diamond Relat. Mater.}} \textbf{17},
  1292--1296 (2008).

\bibitem{Friel2009}
I.~Friel, S.~Clewes, H.~Dhillon, N.~Perkins, D.~Twitchen, and G.~Scarsbrook,
  \enquote{Control of surface and bulk crystalline quality in single crystal
  diamond grown by chemical vapour deposition,} {\protect\JournalTitle{Diam.
  Relat. Mater.}} \textbf{18}, 808--815 (2009).

\bibitem{Tao2014}
Y.~Tao, J.~Boss, B.~Moores, and C.~Degen, \enquote{Single-crystal diamond
  nanomechanical resonators with quality factors exceeding one million,}
  {\protect\JournalTitle{Nat. Commun.}} \textbf{5}, 3638 (2014).

\bibitem{Hwang2004}
D.~Hwang, T.~Saito, and N.~Fujimori, \enquote{New etching process for device
  fabrication using diamond,} {\protect\JournalTitle{Diamond Relat. Mater.}}
  \textbf{13}, 2207 -- 2210 (2004). Proceedings of the 9th International
  Conference on New Diamond Science and Technology (ICNDST-9).

\bibitem{Xie2018}
L.~Xie, T.~X. Zhou, R.~J. St{\"o}hr, and A.~Yacoby, \enquote{Crystallographic
  orientation dependent reactive ion etching in single crystal diamond,}
  {\protect\JournalTitle{Advanced Materials}} \textbf{30}, 1705501 (2018).

\bibitem{Hicks2019}
M.-L. Hicks, A.~C. Pakpour-Tabrizi, V.~Zuerbig, L.~Kirste, C.~Nebel, and R.~B.
  Jackman, \enquote{Optimizing reactive ion etching to remove sub-surface
  polishing damage on diamond,} {\protect\JournalTitle{Journal of Applied
  Physics}} \textbf{125}, 244502 (2019).

\bibitem{Challier2018}
M.~Challier, S.~Sonusen, A.~Barfuss, D.~Rohner, D.~Riedel, J.~Koelbl,
  M.~Ganzhorn, P.~Appel, P.~Maletinsky, and E.~Neu, \enquote{Advanced
  fabrication of single-crystal diamond membranes for quantum technologies,}
  {\protect\JournalTitle{Micromachines}} \textbf{9}, 148 (2018).

\bibitem{Khanaliloo2015}
B.~Khanaliloo, M.~Mitchell, A.~C. Hryciw, and P.~E. Barclay, \enquote{{High-Q/V
  Monolithic Diamond Microdisks Fabricated with Quasi-isotropic Etching},}
  {\protect\JournalTitle{{NANO LETTERS}}} \textbf{{15}}, {5131--5136} ({2015}).

\bibitem{Khanaliloo2015a}
B.~Khanaliloo, H.~Jayakumar, A.~C. Hryciw, D.~P. Lake, H.~Kaviani, and P.~E.
  Barclay, \enquote{Single-crystal diamond nanobeam waveguide optomechanics,}
  {\protect\JournalTitle{Phys. Rev. X}} \textbf{5}, 041051 (2015).

\bibitem{PhysRevLett.116.025001}
Y.~Zhang, C.~Charles, and R.~Boswell, \enquote{Thermodynamic study on plasma
  expansion along a divergent magnetic field,} {\protect\JournalTitle{Phys.
  Rev. Lett.}} \textbf{116}, 025001 (2016).

\bibitem{bazaka_oxygen_2018}
K.~Bazaka, O.~Baranov, U.~Cvelbar, B.~Podgornik, Y.~Wang, S.~Huang, L.~Xu,
  J.~W.~M. Lim, I.~Levchenko, and S.~Xu, \enquote{Oxygen plasmas: a sharp
  chisel and handy trowel for nanofabrication,}
  {\protect\JournalTitle{Nanoscale}} \textbf{10}, 17494--17511 (2018).

\bibitem{SARANGAN2016149}
A.~Sarangan, \enquote{5 - nanofabrication,} in \emph{Fundamentals and
  Applications of Nanophotonics,}  J.~W. Haus, ed. (Woodhead Publishing, 2016),
  pp. 149 -- 184.

\bibitem{Zhang2016a}
Y.~Zhang, C.~Charles, and R.~Boswell, \enquote{Thermodynamic study on plasma
  expansion along a divergent magnetic field,} {\protect\JournalTitle{Phys.
  Rev. Lett.}} \textbf{116}, 025001 (2016).

\bibitem{Bergman1994}
L.~Bergman, M.~McClure, J.~Glass, and R.~Nemanich, \enquote{The origin of the
  broadband luminescence and the effect of nitrogen doping on the optical
  properties of diamond films,} {\protect\JournalTitle{J. Appl. Phys.}}
  \textbf{76}, 3020 (1994).

\bibitem{deOliveira2016}
F.~F. de~Oliveira, S.~A. Momenzadeh, D.~Antonov, J.~Scharpf, C.~Osterkamp,
  B.~Naydenov, F.~Jelezko, A.~Denisenko, and J.~Wrachtrup, \enquote{{Toward
  Optimized Surface delta-Profiles of Nitrogen-Vacancy Centers Activated by
  Helium Irradiation in Diamond},} {\protect\JournalTitle{{NANO LETTERS}}}
  \textbf{{16}}, {2228--2233} ({2016}).

\bibitem{Nelz2019}
R.~Nelz, J.~G{\"o}rlitz, D.~Herrmann, A.~Slablab, M.~Challier, M.~Radtke,
  M.~Fischer, S.~Gsell, M.~Schreck, C.~Becher \emph{et~al.}, \enquote{Toward
  wafer-scale diamond nano-and quantum technologies,}
  {\protect\JournalTitle{APL Materials}} \textbf{7}, 011108 (2019).

\bibitem{Osterkamp2015}
C.~Osterkamp, J.~Lang, J.~Scharpf, C.~M\''uller, L.~P. McGuinness, T.~Diemant,
  R.~J. Behm, B.~Naydenov, and F.~Jelezko, \enquote{Stabilizing shallow color
  centers in diamond created by nitrogen delta-doping using sf6 plasma
  treatment,} {\protect\JournalTitle{Appl. Phys. Lett.}} \textbf{106}, 113109
  (2015).

\bibitem{sangtawesin2018}
S.~Sangtawesin, B.~L. Dwyer, S.~Srinivasan, J.~J. Allred, L.~V. Rodgers,
  K.~De~Greve, A.~Stacey, N.~Dontschuk, K.~M. O'Donnell, D.~Hu \emph{et~al.},
  \enquote{Origins of diamond surface noise probed by correlating single spin
  measurements with surface spectroscopy,} {\protect\JournalTitle{arXiv
  preprint arXiv:1811.00144}}  (2018).

\bibitem{Neu2014}
E.~Neu, P.~Appel, M.~Ganzhorn, J.~Miguel-Sanchez, M.~Lesik, V.~Mille,
  V.~Jacques, A.~Tallaire, J.~Achard, and P.~Maletinsky, \enquote{Photonic
  nano-structures on (111)-oriented diamond,} {\protect\JournalTitle{Appl.
  Phys. Lett.}} \textbf{104}, 153108 (2014).

\bibitem{Radtkenanofab2019}
M.~Radtke, R.~Nelz, L.~Render, and E.~Neu, \enquote{Reliable nanofabrication of
  single-crystal diamond photonic nanostructures for nanoscale sensing,}
  {\protect\JournalTitle{Micromachines}} \textbf{10}, 718 (2019).

\bibitem{Toros2018}
A.~Toros, M.~Kiss, T.~Graziosi, H.~Sattari, P.~Gallo, and N.~Quack,
  \enquote{Precision micro-mechanical components in single crystal diamond by
  deep reactive ion etching,} {\protect\JournalTitle{Microsystems \&
  nanoengineering}} \textbf{4}, 12 (2018).

\bibitem{Pezzagna2010}
S.~Pezzagna, B.~Naydenov, F.~Jelezko, J.~Wrachtrup, and J.~Meijer,
  \enquote{{Creation efficiency of nitrogen-vacancy centres in diamond},}
  {\protect\JournalTitle{New J. Phys.}} \textbf{{12}}, 065017 ({2010}).

\bibitem{Aslam2013}
N.~Aslam, G.~Waldherr, P.~Neumann, F.~Jelezko, and J.~Wrachtrup,
  \enquote{Photo-induced ionization dynamics of the nitrogen vacancy defect in
  diamond investigated by single-shot charge state detection,}
  {\protect\JournalTitle{New J. Phys.}} \textbf{15}, 013064 (2013).

\bibitem{bluvstein2018}
D.~Bluvstein, Z.~Zhang, and A.~C.~B. Jayich, \enquote{Identifying and
  mitigating charge instabilities in shallow diamond nitrogen-vacancy centers,}
  {\protect\JournalTitle{arXiv preprint arXiv:1810.02058}}  (2018).

\bibitem{Shields2015}
B.~J. Shields, Q.~P. Unterreithmeier, N.~P. de~Leon, H.~Park, and M.~D. Lukin,
  \enquote{Efficient readout of a single spin state in diamond via
  spin-to-charge conversion,} {\protect\JournalTitle{Phys. Rev. Lett.}}
  \textbf{114}, 136402 (2015).

\end{thebibliography}

\end{document}